\documentclass[10pt,conference]{IEEEtran}
\IEEEoverridecommandlockouts
\usepackage[T1]{fontenc}
\usepackage{cite}
\usepackage{amsmath,amssymb}
\usepackage{graphicx}
\usepackage{float}
\usepackage{booktabs}
\usepackage{multirow}
\usepackage{tabularx}
\usepackage[most]{tcolorbox}
\usepackage[hidelinks]{hyperref}
\hypersetup{
  pdftitle={AiXapply: Fast and Reliable Full-File Code Integration with Specialized Small Models for IDE Workflows},
  pdfauthor={Siyuan Jiang, Xiang Cai, Peixu Wang, Yu Han, Yihong Dong, Wei Ning, Xuyuan Guo, Jincheng Wen, Dexin Liu, Wei Zhao, Ge Li},
  pdfkeywords={Code Integration, Large Language Model, Reinforcement Learning}
}

\title{AiXapply: Fast and Reliable Full-File Code Integration with Specialized Small Models for IDE Workflows\thanks{Project repository: \url{https://github.com/aixcoder-plugin/aiXapply-4B}.}}
\author{
\IEEEauthorblockN{Siyuan Jiang\IEEEauthorrefmark{1}\IEEEauthorrefmark{3},
Xiang Cai\IEEEauthorrefmark{1}\IEEEauthorrefmark{3},
Peixu Wang\IEEEauthorrefmark{2},
Yu Han\IEEEauthorrefmark{1},
Yihong Dong\IEEEauthorrefmark{2},\\
Wei Ning\IEEEauthorrefmark{1},
Xuyuan Guo\IEEEauthorrefmark{2},
Jincheng Wen\IEEEauthorrefmark{2},
Dexin Liu\IEEEauthorrefmark{1},
Wei Zhao\IEEEauthorrefmark{2},
Ge Li\IEEEauthorrefmark{2}\IEEEauthorrefmark{4}}
\IEEEauthorblockA{\IEEEauthorrefmark{1}aiXcoder \qquad \IEEEauthorrefmark{2}Peking University}
\IEEEauthorblockA{\{jiangsiyuan,caixiang,hangyu,ningwei\}@aixcoder.com, ldx@aixcoder.com\\
peixu.wang@u.nus.edu, dongyh@stu.pku.edu.cn, xyguo25@stu.pku.edu.cn\\
3116127286@qq.com, \{zhaowei,lige\}@pku.edu.cn}
\thanks{\IEEEauthorrefmark{3}These authors contributed equally to this work. \IEEEauthorrefmark{4}Ge Li is the corresponding author.}
}

\begin{document}
\maketitle
\begin{abstract}
  Large language models increasingly assist developers by proposing small code edits, yet reliably integrating a partial update snippet into an existing file while preserving unrelated content remains poorly supported by current code-generation workflows.
 We formalize the Code Integration (Apply) task as follows: given an original file and an update snippet, the system must produce the complete merged file while preserving all content outside the requested edit. We construct a high-quality multi-language dataset from real commits via a structured pipeline that converts code changes into update snippets paired with full-file ground truth and enforces consistency through verification and filtering. We then train a specialized 4B model, \textbf{aiXapply}, using supervised fine-tuning and reinforcement learning (GRPO). The supervised variant achieves 94.4\% accuracy under our language-aware comparison, surpassing a strong large-model (Deepseek-V3.2) baseline (91.6\%). Speculative decoding enables aiXapply-RL to reach $\sim$2700 tokens/s on a single A100 in a decoding subset; on the full benchmark, it achieves roughly 75$\times$ lower latency than the API-served Deepseek-V3.2 baseline. We further conduct an industrial deployment of aiXapply-RL in the aiXcoder IDE plugin as a dedicated Apply service, reducing mean Apply latency from 50 seconds to 1.89 seconds by offloading full-file generation from the upstream large model. Under time-based capacity planning, replacing one of five upstream serving instances would save 20\% of upstream machine cost before the additional shared A100 service cost. Overall, our results demonstrate that treating code integration as a standalone task enables small models to achieve high accuracy, practical efficiency, and concrete industrial benefits in real-world IDE workflows where latency, cost and privacy are critical.
\end{abstract}

\section{Introduction}
AI-assisted IDE workflows often generate localized \emph{update snippets} that must be merged into a source file while preserving unrelated content. In aiXcoder, using the upstream model for full-file Apply took 50 seconds on average within an approximately 240-second editing workflow. We isolate this integration stage as the Apply task and offload it to a specialized small model.

Unified diff represents edits through hunk headers, line-level changes, and surrounding context. Small errors in headers or context alignment can make a generated patch non-applicable.

Search-and-replace avoids diff syntax but requires the intended source fragment to be identified. Unmatched or ambiguous fragments can cause failure, so it is commonly used with iterative retries rather than for one-shot integration.

\begin{figure}[!t]
    \centering
    \includegraphics[width=1\linewidth]{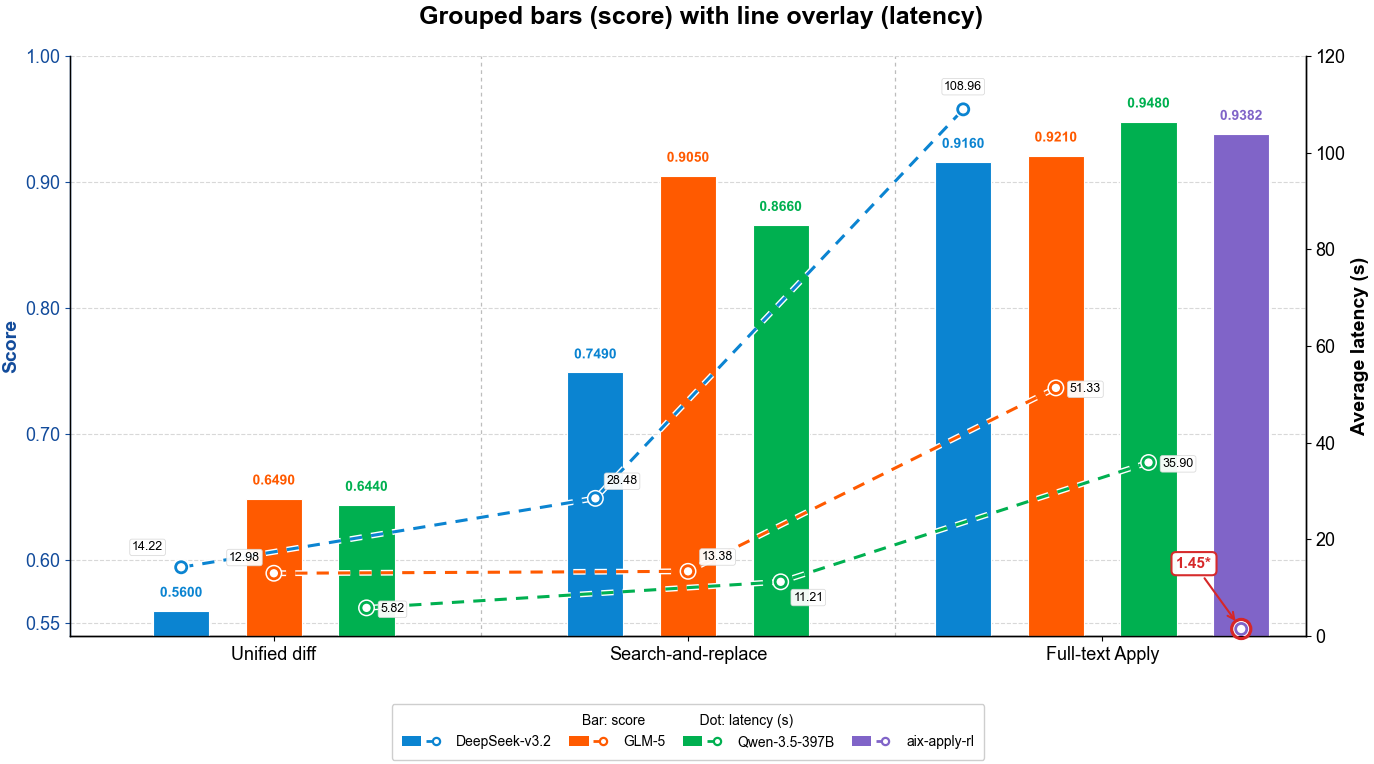}
    \caption{
Accuracy–latency comparison across three code integration paradigms. Bars show equivalence accuracy for different models, while the line indicates average latency (seconds). For DeepSeek-V3.2, moving from unified diff to search-and-replace and then to full-file Apply consistently improves accuracy (0.560 $\rightarrow$ 0.749 $\rightarrow$ 0.916), but also increases latency (14.22s $\rightarrow$ 28.48s $\rightarrow$ 108.96s). Across models, full-file Apply achieves the highest accuracy, with Qwen3.5-397B reaching 0.948. Our aiXapply-RL maintains comparable accuracy (0.938) while reducing latency to 1.45s, demonstrating a significantly better accuracy–latency trade-off for one-shot code integration.
}
    \label{fig:main result}
\end{figure}

\begin{figure*}[!t]
    \centering
    \includegraphics[width=\textwidth]{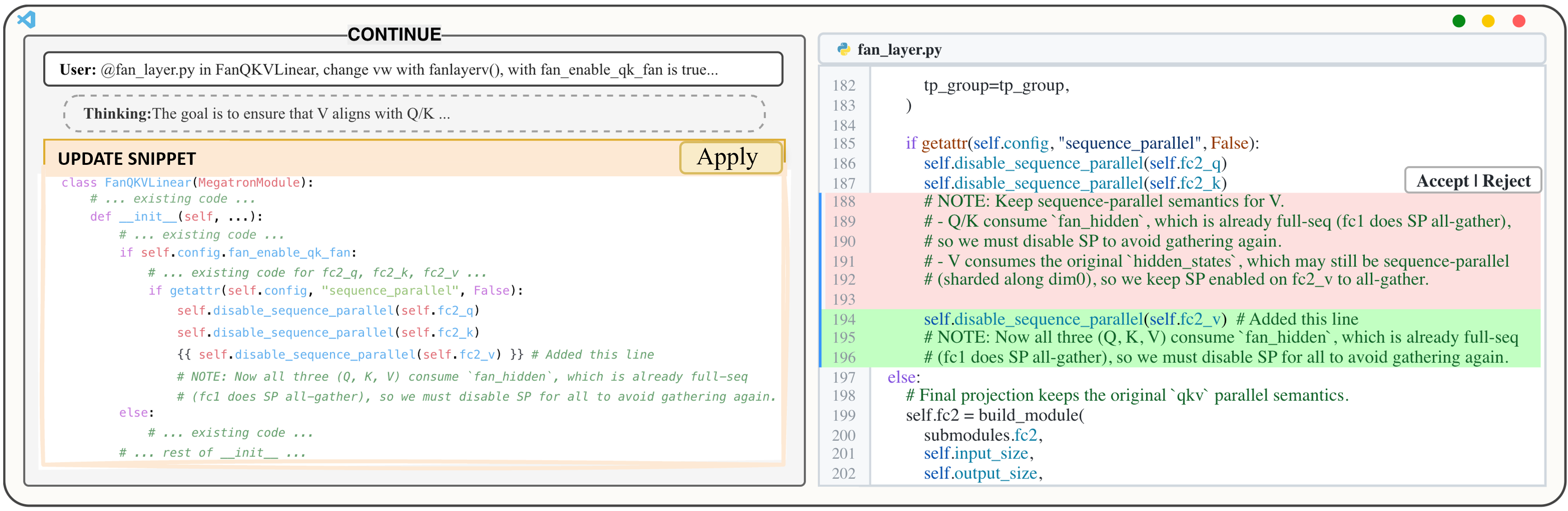}
    \caption{Workflow of aiXapply in VS Code. A user submits an editing request through Continue, the LLM then generates an update snippet, which aiXapply applies to the source file to produce a candidate updated file. VS Code displays the resulting diff for user inspection, after which the user decides whether to accept or reject the proposed edit.}
    \label{fig:task_intro}
\end{figure*}

Full-file Apply generates the complete updated file directly, avoiding intermediate representations. However, regenerating unchanged content with a large model incurs substantial latency and token cost.

This gap motivates a different perspective.
Rather than treating editing as a byproduct of general code generation, we formulate snippet integration as a dedicated software engineering task: given a source file and an update snippet, produce the merged file with zero unintended changes outside the target region.
This formulation is precise, verifiable, and directly aligned with IDE requirements.

In this workflow, the upstream model specifies the intended change as a snippet, while a dedicated Apply model handles full-file reproduction. This offloads the copying of unchanged content from the large model to a smaller model, whose copy-dominant workload can be accelerated with speculative decoding.

Our goal is to retain the accuracy of full-file Apply while making it efficient enough for deployment. We study this problem through four research questions.

\textbf{RQ1: What are the advantages and limitations of full-file Apply compared to unified diff and search-and-replace?}

\textbf{RQ2: Can a small model achieve high accuracy and robustness on the Apply task across diverse languages and out-of-distribution scenarios?}

\textbf{RQ3: Can full-file Apply be made efficient for realistic IDE scenarios?}

\textbf{RQ4: What training strategies are effective for the Apply task?}

We compare the three integration paradigms, train 4B SFT and RL variants on commit-derived data, and evaluate accuracy, robustness, and decoding efficiency across 20 languages. We also report the aiXcoder deployment and its manual Apply assessment on real development tasks. Figure~\ref{fig:main result} summarizes the accuracy--latency trade-off. Our contributions are:

\begin{itemize}
    \item \textbf{Industrial deployment and practical evidence:}
    We integrate a dedicated Apply service into aiXcoder, report its latency reduction, and evaluate first-output Apply acceptability on 100 real development tasks (91 acceptable with aiXapply-RL versus 88 with GLM-5).

    \item \textbf{A systematic evaluation of code integration paradigms:}
    We compare full-file Apply with search-and-replace and unified diff across accuracy, efficiency, and generalization benchmarks, and analyze SFT/RL trade-offs across languages and challenging settings.

    \item \textbf{An accurate and efficient small model:}
    We provide \textbf{aiXapply}, a specialized 4B model trained with supervised fine-tuning and reinforcement learning, achieving high equivalence accuracy while enabling fast inference with speculative decoding.

    \item \textbf{A high-quality dataset for code Apply:}
    We provide a multi-language dataset from real-world commits via a structured pipeline that converts code changes into update snippets paired with full-file ground truth, with strict verification to ensure correctness and reproducibility.

\end{itemize}

\section{Related Work}

Large language models have been widely applied to code editing tasks conditioned on natural language instructions or surrounding context \cite{chen2021evaluating, cassano2023can, li2024instructcoder}. Benchmarks such as CodeEditorBench \cite{guo2024codeeditorbench} and EDIT-Bench \cite{chi2025edit} evaluate how well models follow editing instructions. Unlike these efforts, our work assumes an update snippet has already been provided and focuses on the downstream integration step: faithfully merging that snippet into the original file while preserving all unchanged content. This shifts the emphasis from creative generation to deterministic, low-latency integration with zero side effects.

The choice of edit representation critically affects reliability. Unified diffs \cite{le2011genprog} are precise but brittle; search-and-replace formats \cite{fastapply_github} depend on source-fragment matching. Industrial tools like Aider \cite{aider_chat} and Cursor's Instant Apply \cite{cursor_instant_apply} have adopted various strategies and even trained specialized apply models. Most related to our work is FuseApplyBench \cite{liang2025fuseapplybench}, which benchmarks trustworthy code edit applying, and the FastApply pipeline \cite{fastapply_github}. Diff-XYZ \cite{glukhov2025diff} evaluates diff understanding. Our contribution is an open 4B implementation, a commit-derived dataset, and empirical evidence on editing representations, training trade-offs, and enterprise deployment.

Learning from version control history is a long-standing approach. CommitPack \cite{muennighoff2023octopack} provides large-scale real-world commits; CodeEditor \cite{li2023codeeditor} and Coeditor \cite{wei2024coeditor} learn edit patterns from diffs. However, these methods focus on generating the edit itself, whereas we tackle the distinct problem of integration. Furthermore, end-to-end software engineering agents (e.g., SWE-bench \cite{jimenez2023swe}, SWE-agent \cite{yang2024swe}, Agentless \cite{xia2024agentless}) address the full pipeline from issue localization to patch generation. Our work complements them by isolating and optimizing the final, often brittle, apply step—a necessary component for any agent that must operate safely in an IDE.

Finally, efficient inference is crucial for interactive use. Speculative decoding \cite{leviathan2023fast, chen2023accelerating} accelerates general generation, while methods such as REST \cite{he2024rest} and Medusa \cite{cai2024medusa} further improve decoding efficiency. More recently, EfficientEdit \cite{wang2025efficientedit} leverages the copy-dominant nature of code editing to accelerate editing-specific workloads. In contrast, our work focuses on the full-file Apply setting, where most tokens are preserved from the source file, and explores lightweight proposal strategies tailored to this structured editing scenario.

\section{Data Construction}

We construct training examples for the Apply task, where each example is a triple \texttt{<source\_file, update\_snippet, update\_file>}
that satisfies a strong causality requirement: the full update must be recoverable from the snippet alone. To do so, we start from real-world commits and convert them into Apply-style examples through a multi-stage pipeline. The pipeline is designed to preserve realistic edit intent while filtering out ambiguous, low-quality, or non-reproducible samples.
\subsection{Task Definition}

\paragraph{Formalization}
We define the Apply task as follows. Given:
\begin{equation*}
x = (L, S, U),
\end{equation*}
where $L$ is the programming language or file type, $S$ is the original full file, and $U$ is an update snippet, the goal is to produce:
\begin{equation*}
Y = S',
\end{equation*}
the updated full file.

We adopt a structured I/O format to reduce ambiguity:
\begin{itemize}
    \item \textbf{Input:}
\begin{verbatim}
<language>{L}</language>
<source_file>{S}</source_file>
<update_snippet>{U}</update_snippet>
\end{verbatim}
    \item \textbf{Output:}
\begin{verbatim}
<update_file>{S'}</update_file>
\end{verbatim}
\end{itemize}

\paragraph{Core constraints}
The task has two target constraints:
\begin{itemize}
    \item \textbf{Non-side-effect constraint:} Outside the patch region, $S'$ must be \emph{identical} to $S$ at the lexical level (including comments, ordering).
    \item \textbf{Deterministic full-file output:} The model must output a complete file that can be directly written back to disk, without omissions or explanatory text.
\end{itemize}

\paragraph{Update snippet semantics}
The update snippet $U$ is a partial view of the modified code, with the following semantics:
\begin{itemize}
    \item \textbf{Placeholders:} Tokens such as \texttt{// ... existing code ...} indicate that the corresponding content must be copied \emph{exactly} from $S$ into $S'$, and must not appear in the final output.
    \item \textbf{Deletion:} Deletions can be expressed implicitly by omitting blocks or via explicit markers (e.g., \texttt{// remove ...}).
    \item \textbf{Anchor uniqueness:} Context anchors extracted from $U$ must match a \emph{unique} location in $S$. If no unique match exists, the model should fail safely.
\end{itemize}

\paragraph{Decomposition of correctness}
We decompose Apply into three verifiable steps:
\begin{enumerate}
    \item \textbf{Context Matching:} Locate a unique region in $S$ using anchors from $U$;
    \item \textbf{Code Merging:} Replace the corresponding region with updated content from $U$;
    \item \textbf{Expansion:} Resolve placeholders by copying exact content from $S$.
\end{enumerate}

\begin{figure}[t]
    \centering
    \includegraphics[width=\linewidth]{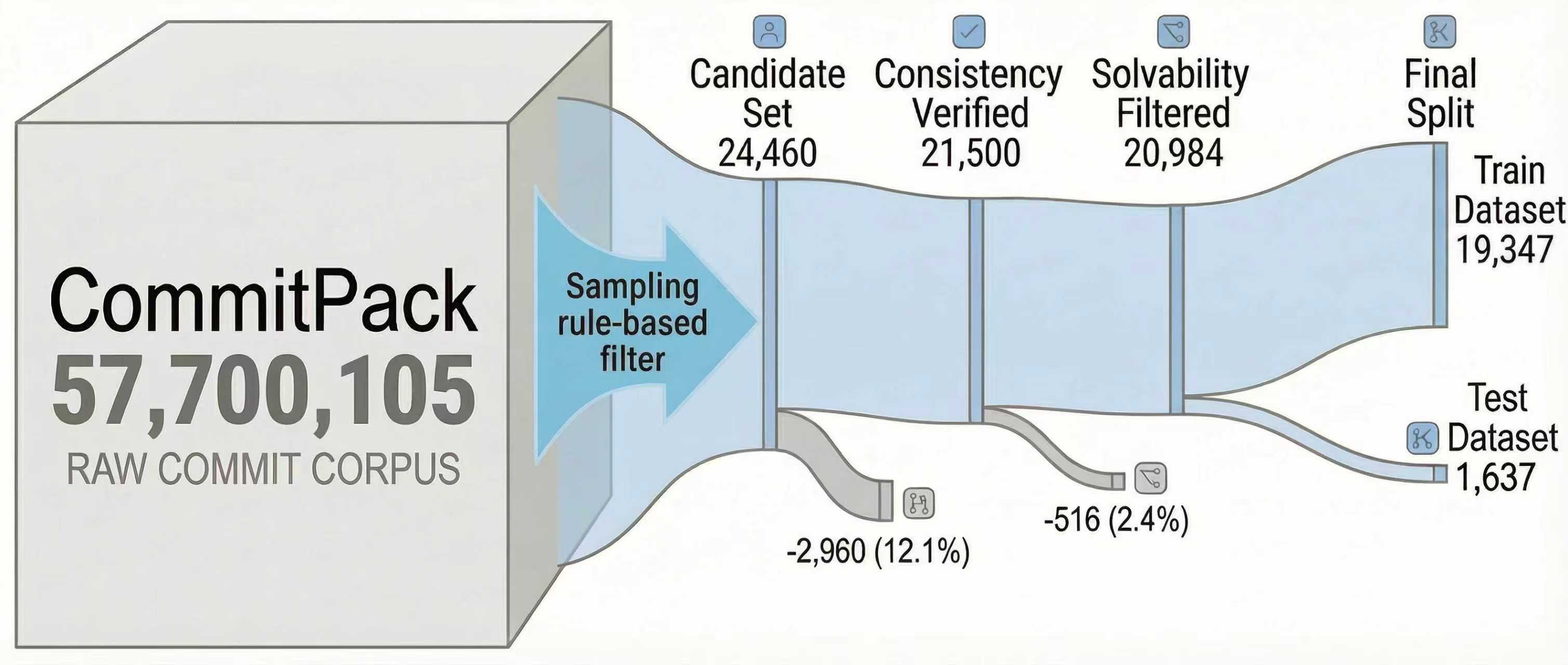}
    \caption{
    Overview of the dataset construction pipeline. Starting from 57.7M raw commits in CommitPack, rule-based sampling produces 24,460 candidate samples. Consistency verification removes 12.1\% of samples, followed by solvability filtering that removes an additional 2.4\%, resulting in 20,984 high-quality examples. The final dataset is split into 19,347 training samples and 1,637 test samples.
    }
    \label{fig:dataset_pipeline}
\end{figure}

\subsection{Dataset pipeline}

To construct training data that satisfies the task constraints, we design a multi-stage pipeline that converts raw commits into well-formed Apply examples. Starting from real-world version control data, the pipeline progressively refines each sample by making edit intent explicit, enforcing locality and determinism, and filtering out ambiguous or non-reproducible cases. Figure~\ref{fig:dataset_pipeline} illustrates the overall data flow and sample reduction across stages.

\subsubsection{Source Data and Sampling}

We use the \texttt{CommitPack} dataset from HuggingFace as the data source, which provides real-world code edits extracted from version control histories. Each record contains the file content before and after a commit (\texttt{old\_contents}, \texttt{new\_contents}) along with commit metadata, enabling direct supervision of code changes at the file level.

To align the data with the snippet integration task, we apply rule-based filtering and sampling. We focus on single-file edits and control file length, edit size, and line-change ratio to ensure localized modifications. We also filter commits based on basic quality criteria (e.g., commit message length) and perform language-balanced sampling to ensure broad coverage across programming languages and structured file types. After this stage, the dataset is reduced from 57.7M raw commits to 24,460 high-quality candidate samples.

\subsubsection{From Commit to Explicit Change Description}

A raw commit provides before/after code and a short commit message, but this is not yet an Apply example. In the first generation stage, we use an LLM to convert each commit into an explicit \texttt{change\_description}. The input includes the language, original code, changed code, and commit message, and the output is a concise description of what changed, where it changed, and how the change should be interpreted.

This stage serves two purposes. First, it makes the edit intent more explicit than a short commit message. Second, it simulates the upstream setting in which another model or developer has already identified the desired modification before the Apply model is invoked. The description is treated as an intermediate specification rather than final supervision.

\subsubsection{Synthesizing the Update Snippet and Ground Truth}

Given the original file and the generated description, we use a second LLM stage to produce a training pair: an \texttt{update\_snippet} and the corresponding full \texttt{final\_code}.

The \texttt{update\_snippet} is a localized code fragment centered on the modified region. It contains updated code near the change location, while unchanged content outside the region is replaced with placeholders such as \texttt{// ... existing code ...}. These placeholders preserve sufficient context for locating the correct position in the original file. For deletions, the snippet retains surrounding code and omits the removed block.

The \texttt{final\_code} is the complete ground-truth file obtained by applying the snippet semantics to the original file. It must contain the entire file with no placeholders and reflect only the intended change. Unchanged regions must remain identical at the lexical level, including formatting and comments.

This stage converts commit-level supervision into the exact input-output format required by the Apply task: a localized snippet paired with its deterministic full-file result.
The editing intent is grounded in real commits; synthesis emulates the partial-update snippets produced by upstream coding models. The target is the full file obtained by applying the generated snippet, not necessarily the original commit's after-file. Consistency verification checks that the snippet explains the resulting file changes.

\subsubsection{Consistency Verification}

A generated snippet is useful only if it fully explains the resulting file change. We therefore introduce a verification stage that checks whether the pair (\texttt{update\_snippet}, \texttt{final\_code})
is causally consistent with the original file. The LLM-based verifier receives the original code, the final code, the deterministic diff between them, and the snippet, and judges whether the diff can be explained completely and only by the snippet. Samples that fail this check are discarded.

This stage checks for strict causality: every change in the diff must be supported by the snippet, and the snippet must not imply additional changes. Placeholders are treated as keep-original directives rather than free-form omissions. This step removes 2,960 samples (12.1\%), resulting in 21,500 consistency-verified examples.

\subsubsection{Solvability Filtering}

After consistency verification, we further filter examples based on solvability. For each sample, Qwen3.5-397B-A17B, DeepSeek-V3.2, and GLM-5 each perform the Apply task twice, giving six attempts from
\(
(\texttt{source\_file}, \texttt{update\_snippet})
\).
Each predicted full file is compared with the ground truth using token-level equivalence.

The failure-count distribution showed a sharp drop in the number of samples with five or six failures. Manual spot checks found data-quality problems in most inspected samples from these groups, so we discard samples that fail five or all six attempts. This step removes 516 samples (2.4\%), yielding 20,984 retained examples.
This model-dependent filter may also exclude valid but difficult edits, biasing the retained samples toward cases that these models can solve; it does not require every model or attempt to succeed.

Finally, the dataset is randomly split at the commit level into 19,347 training samples and 1,637 test samples, forming a high-quality benchmark for the Apply task.

\subsection{Error Taxonomy}

To support diagnosis and training, we categorize failures into six types:

\begin{itemize}
    \item \textbf{OUTPUT\_INVALID:} Invalid format (e.g., missing wrapper, truncation, placeholder leakage, or unparsable structure).
    \item \textbf{PATCH\_NOT\_APPLIED:} No effective change applied.
    \item \textbf{PATCH\_INCOMPLETE:} Partial update with missing lines or blocks.
    \item \textbf{PATCH\_INCORRECT:} Incorrect modification within the patch region.
    \item \textbf{WRONG\_POSITION:} Correct content applied to the wrong location.
    \item \textbf{OUT\_OF\_PATCH\_SIDE\_EFFECT:} Unintended modifications outside the patch region.
\end{itemize}

This taxonomy captures both hard failures (non-writable outputs) and semantic errors (misplacement, incompleteness, side effects), enabling fine-grained analysis and alignment.

\section{Method}

We study full-file Apply as a deterministic code integration task and explore two training paradigms: supervised fine-tuning and reinforcement learning.
Both approaches are trained independently.
At inference time, we further improve efficiency using speculative decoding.

\subsection{Model and Prompting}

We conduct training on the Qwen3-4B \cite{yang2025qwen3} model with a maximum context length of 64k.
The model is trained as a \textit{deterministic patching engine}: given an original file and an update snippet, it directly generates the final merged file.

Each input consists of:
(1) the original file (\texttt{old\_contents}), and
(2) the update snippet (\texttt{update\_snippet}).
The prompt instructs the model to apply the snippet to the original file.

\subsection{Supervised Fine-tuning}

The SFT model is trained with standard next-token cross-entropy over the complete target file. Supervising unchanged regions encourages identity copying and discourages unintended modifications.
The released SFT configuration (artifact revision \texttt{24c4e3}) defaults to a learning rate of $10^{-5}$, a per-device batch size of 1, 64 gradient-accumulation steps, and two epochs.

For long-context training, we concatenate the structured input (\texttt{source\_file}, \texttt{update\_snippet}) with the target \texttt{update\_file} and truncate sequences to the maximum length (32k tokens for the concatenated sequence in the released SFT configuration). No partial loss masking is applied on the output, ensuring that every generated token contributes to the training signal.

\subsection{Reinforcement Learning}

We train a separate model using Group Relative Policy Optimization (GRPO) \cite{shao2024deepseekmath}. For each input, sampled outputs are scored using the rewards below, and group-normalized advantages drive a clipped policy objective with KL regularization to a reference policy. This allows task-level penalties for incomplete edits and side effects without a separate value function.
The released GRPO configuration defaults to a learning rate of $5\times10^{-7}$, a training batch of 128 prompts, eight rollouts per prompt, a KL-loss coefficient of 0.01, and three epochs.

\subsection{Reward Design}
\label{sec:reward-design}

We design a reward function that directly reflects the core requirement of the Apply task: \textbf{exact integration with zero unintended side effects}. Unlike general code generation, Apply admits a deterministic notion of correctness—namely, whether the generated full file is \emph{equivalent} to the ground truth under a precise lexical or structural criterion.

\paragraph{Equivalence-Based Reward.}
Given a model prediction $\hat{y}$ and ground truth $y$, we define a binary reward:
\begin{equation}
r(\hat{y}, y) =
\begin{cases}
1, & \text{if } \hat{y} \equiv y \\
0, & \text{otherwise}
\end{cases}
\end{equation}

The equivalence relation $\equiv$ is implemented via a language-aware comparison:

\begin{itemize}
    \item \textbf{Code files:} We apply token-level equivalence using lexical analysis (Pygments). Both $\hat{y}$ and $y$ are tokenized, ignoring whitespace-only tokens, and compared for exact match at the token sequence level.
    \item \textbf{Structured formats (e.g., JSON, YAML, XML):} We parse both outputs into canonical structures and compare their semantic representations.
    \item \textbf{Text-like formats (e.g., Markdown, SQL):} We apply the format-specific normalization in the released evaluator before exact string comparison.
\end{itemize}

This design tolerates formatting differences while requiring exact agreement under the above language-aware comparison rules.

\paragraph{Error-Aware Dense Reward.}
While the binary reward provides a clear correctness signal, it is sparse and does not distinguish between different failure modes.
To provide more informative feedback, we introduce a dense reward that explicitly accounts for error types.

Given a prediction that is not equivalent to the ground truth, we analyze two types of errors:
(1) \emph{patch errors}, defined as the number of incorrect lines within the target modification region, and
(2) \emph{side effects}, defined as the number of unintended modified lines outside the target region.

The reward is defined as:
\begin{equation}
r(\hat{y}, y) = \max \left(0,\; 1 - \alpha \cdot \text{patch\_errors} - \beta \cdot \text{side\_effects} \right),
\end{equation}
where $\alpha=0.25$ and $\beta=0.35$ in the dense reward used by the released and deployed RL model, with a larger penalty assigned to side effects.

This design emphasizes a key principle of the Apply task: errors outside the target region are more harmful than imperfect edits within the patch.
By distinguishing error types and measuring them at the line level, the reward provides finer-grained guidance and encourages models to produce correct edits while preserving unchanged regions.

\subsection{Speculative Decoding}

Full-file Apply is copy-dominant. We use speculative decoding to propose token blocks and verify them in parallel \cite{leviathan2023fast}, comparing the following model-free proposal mechanisms.

\textbf{N-gram-based.}  
We first employ an n-gram prompt-lookup strategy as a lightweight proposal mechanism. An n-gram dictionary is constructed from the prompt and previously generated tokens, mapping fixed-length suffixes to observed continuations. At each decoding step, the current suffix is used to retrieve candidate continuations, which are merged into a prefix tree and verified in a single forward pass. This approach is efficient and works well when local token patterns repeat frequently.

\textbf{Suffix-based.}  
In parallel, we employ a suffix-based, model-free speculative mechanism that indexes previously generated token sequences in a suffix structure and, at each step, retrieves continuations by finding the longest suffix match of the current context. Longer matches imply higher confidence, so the method adaptively proposes longer candidate continuations when strong matches are found and shorter ones otherwise. These candidates are organized into a small, frequency-prioritized speculation tree and verified in a single forward pass of the target model, enabling efficient multi-token validation without relying on auxiliary draft models.

\textbf{Evaluation Metrics.}  
We report mean and P95 request latency, and throughput as generated tokens divided by total execution time. The inference-speed experiments use aiXapply-RL with a single concurrent request, measuring latency end to end for each request.

\section{Experiments}

\subsection{Experimental Setup}

\subsubsection{Task, Metrics, and Baselines}

For full-file Apply, all models use the same structured input format:
\texttt{<language> + <source file> + <update snippet> -> <updated file>}.
The target output is always the full merged file.
For RQ1, search-and-replace edits are executed by the OpenCode-style tool, including its whitespace- and indentation-tolerant matching, and unified diffs by our custom executor. Both implementations are released with our artifact. All three paradigms use the same language-aware correctness comparison on the resulting full file.
\emph{Equivalence accuracy} (Avg/Overall) is total correct predictions divided by total samples; code files use Pygments-token equivalence, while structured and text-like formats use parsing when appropriate.
An output is counted as correct when it matches the reference under the rules in Section~\ref{sec:reward-design}, rather than requiring byte-for-byte equality. Normalization can hide meaningful changes, and behavior-preserving rewrites may fail the comparison.

We compare against:
(1) Qwen3-4B (4B base model),
(2) DeepSeek-V3.2 (strong large-model baseline \cite{liu2025deepseek}),
(3) Qwen3.5-397B-A17B (strong large-model baseline \cite{qwen3.5}),
(4) GLM-5 (strong large-model baseline \cite{zeng2026glm}),
(5) Fast-Apply-7B (a recent code apply model baseline \cite{fastapply_github}),
(6) aiXapply-SFT (our supervised model),
(7) aiXapply-RL (our RL-aligned model)
.

\subsubsection{Datasets}

The main aiXapply test set contains 1,637 examples covering 20 languages / file types.
The groups are Java and JavaScript (200 each), Python (195), C (130), C++ (128), Go (80), and JSON (54), with 50 examples in each of the remaining 13 languages / file types.
To evaluate robustness, we construct additional generalization dataset which include Long Context, Random Placeholders, Chunk File, and Untrained Languages covering PHP, CSS, SystemVerilog, and CSharp.
The Long Context and Untrained Languages sets contain 51 and 647 examples, respectively. We also report transfer results on Diff-XYZ \cite{glukhov2025diff}.

\subsection{RQ1: Advantages and Limitations of Full-File Apply}

We first compare three code integration paradigms: unified diff, search-and-replace, and full-file Apply. Table~\ref{tab:rq1_overall} shows that full-file Apply consistently achieves the highest accuracy. On DeepSeek-v3.2, accuracy improves from 0.560 (diff) and 0.749 (search-replace) to 0.916 (Apply), a gain of 16.7 percentage points over search-and-replace. This trend is consistent across all models, indicating that the advantage mainly comes from the representation rather than model scale.

The weakness of unified diff may lie in structural brittleness. Our qualitative analysis identified malformed diff headers and incorrect hunk line numbers or line counts among the observed failures. Its strict format requirements make it sensitive to small errors in hunk headers or context alignment, which can render the entire patch invalid. These failures are largely independent of the actual edit semantics.

Search-and-replace improves robustness by removing patch syntax, but it still relies on matching source fragments. Even with whitespace and indentation tolerance, unmatched or ambiguous fragments can cause replacement failure, limiting its reliability in one-shot generation. This makes it more suitable for iterative agent workflows than for deterministic integration.

\begin{table}[!htbp]
\centering
\caption{Overall accuracy comparison of three code integration paradigms.}
\label{tab:rq1_overall}
\small
\setlength{\tabcolsep}{3pt}
\begin{tabularx}{\columnwidth}{lccc}
\toprule
Model & Unified diff & Search-and-replace & Full-file Apply \\
\midrule
DeepSeek-v3.2 & 0.560 & 0.749 & \textbf{0.916} \\
GLM-5 & 0.649 & 0.905 & \textbf{0.921} \\
Qwen-3.5-397B & 0.644 & 0.866 & \textbf{0.948} \\
\bottomrule
\end{tabularx}
\end{table}

\begin{table}[!htbp]
\centering
\caption{Average output length and latency across the three paradigms.}
\label{tab:rq1_efficiency}
\small
\setlength{\tabcolsep}{4pt}

\resizebox{\columnwidth}{!}{
\begin{tabular}{lcccccc}
\toprule
& \multicolumn{2}{c}{Diff} 
& \multicolumn{2}{c}{Search \& Replace} 
& \multicolumn{2}{c}{Apply} \\
\cmidrule(lr){2-3} \cmidrule(lr){4-5} \cmidrule(lr){6-7}
Model 
& Tokens & Latency (s)
& Tokens & Latency (s)
& Tokens & Latency (s) \\
\midrule
DeepSeek-v3.2 & 462 & 14.22 & 926 & 28.48 & 3684 & 108.96 \\
GLM-5 & 437 & 12.98 & 512 & 13.38 & 3646 & 51.33 \\
Qwen-3.5-397B-A17B & 460 & 5.82 & 902 & 11.21 & 3674 & 35.90 \\
aiXapply-RL & -- & -- & -- & -- & 3734 & 1.45 \\
\bottomrule
\end{tabular}
}

\end{table}

Full-file Apply achieves the best performance by eliminating both syntax and matching constraints. The model only needs to generate the final file, which aligns naturally with autoregressive training and leads to consistently higher accuracy.

The main limitation of full-file Apply is efficiency. As shown in Table~\ref{tab:rq1_efficiency}, it requires around 3684 tokens and 108.96 seconds on DeepSeek-v3.2, compared to 462 tokens and 14.22 seconds for unified diff. This overhead comes from regenerating unchanged content.

This cost is largely due to the copy-dominant nature of the task. By leveraging speculative decoding, our system reduces latency to 1.45 seconds while maintaining the same output length, showing that the efficiency gap can be largely mitigated.

Overall, unified diff is compact but too brittle for direct generation. Search-and-replace improves robustness but still depends on source-fragment matching. Full-file Apply provides the highest one-shot accuracy by removing fragile intermediate representations, with higher decoding cost that can be effectively optimized in practice.

\begin{table*}[!htbp]
\centering
\caption{Main results accuracy on the aiXapply test dataset.}
\label{tab:main-benchmark}

\small
\setlength{\tabcolsep}{3pt}

\resizebox{\textwidth}{!}{
\begin{tabular}{lccccccccccccccccccccccc}
\toprule
\textbf{Model/Language} 
& \textbf{C} & \textbf{C++} & \textbf{Docker} & \textbf{Go} & \textbf{HTML} & \textbf{INI} 
& \textbf{Java} & \textbf{JS} & \textbf{JSON} & \textbf{Make} & \textbf{MD} 
& \textbf{Python} & \textbf{RST} & \textbf{Rust} & \textbf{Shell} 
& \textbf{SQL} & \textbf{Text} & \textbf{TS} & \textbf{XML} & \textbf{YAML} 
& \textbf{Avg} \\
\midrule

Qwen3-4B 
& 0.66 & 0.66 & 0.68 & 0.65 & 0.64 & 0.50 & 0.71 & 0.71 & 0.81 & 0.62 & 0.54 & 0.69 & 0.52 & 0.48 & 0.58 & 0.60 & 0.54 & 0.62 & 0.76 & 0.54 & 0.651 \\

DeepSeek-v3.2 
& 0.915 & 0.914 & 0.94 & 0.937 & 0.84 & 0.88 & 0.955 & 0.934 & 0.944 & 0.92 & 0.82 & 0.928 & 0.88 & 0.90 & 0.92 & 0.84 & 0.90 & 0.96 & 0.86 & 0.92 & 0.916 \\

Qwen3.5-397B-A17B
& 0.946 & \textbf{0.977} & \textbf{1.00} & 0.95 & \textbf{0.96} & 0.94 & 0.945 & 0.96 & \textbf{1.00} & \textbf{0.98} & 0.90 & 0.933 & 0.86 & \textbf{0.90} & \textbf{0.98} & 0.94 & 0.88 & 0.92 & \textbf{1.00} & 0.96 & \textbf{0.948} \\

GLM-5
& 0.931 & 0.945 & 0.96 & 0.963 & 0.92 & \textbf{0.98} & 0.95 & 0.93 & 0.944 & \textbf{0.98} & 0.88 & 0.897 & 0.86 & 0.82 & 0.92 & 0.92 & 0.86 & 0.94 & 0.78 & 0.90 & 0.921 \\

Fast-Apply-7B
& 0.546 & 0.539 & 0.74 & 0.65 & 0.60 & 0.64 & 0.545 & 0.73 & 0.814 & 0.56 & 0.66 & 0.626 & 0.64 & 0.46 & 0.66 & 0.60 & 0.64 & 0.48 & 0.62 & 0.76 & 0.620 \\

aiXapply-RL 
& \textbf{0.96} & 0.92 & \textbf{1.00} & 0.93 & 0.86 & 0.92 & 0.95 & \textbf{0.97} & 0.96 & 0.92 & \textbf{0.90} & 0.91 & 0.88 & \textbf{0.90} & \textbf{0.98} & \textbf{0.96} & \textbf{1.00} & 0.88 & 0.98 & 0.96 & 0.938 \\

aiXapply-SFT
& 0.93 & 0.945 & \textbf{1.00} & 0.912 & \textbf{0.96} & 0.96 & 0.93 & 0.965 & \textbf{1.00} & 0.94 & 0.88 & \textbf{0.933} & \textbf{0.90} & \textbf{0.90} & 0.92 & \textbf{0.96} & 0.96 & \textbf{0.98} & 0.98 & \textbf{0.98} & 0.944 \\

\bottomrule
\end{tabular}
}

\end{table*}

\begin{table}[!t]
\centering
\caption{Accuracy on Long Context setting.}
\label{tab:long-context}
\begin{tabular}{lc}
\toprule
\textbf{Model} & \textbf{Accuracy} \\
\midrule
Qwen3-4B-Baseline & 0.2353 \\
DeepSeek-V3.2 & 0.588 \\
aiXapply-RL & 0.6471 \\
aiXapply-SFT & \textbf{0.843} \\
\bottomrule
\end{tabular}
\end{table}

\begin{table}[!t]
\centering
\caption{Accuracy on untrained programming languages.}
\label{tab:untrained}
\begin{tabular}{lccccc}
\toprule
\textbf{Model} & \textbf{C\#} & \textbf{CSS} & \textbf{PHP} & \textbf{SV} & \textbf{Avg} \\
\midrule
Qwen & 0.6356 & 0.623 & 0.6409 & 0.6695 & 0.6399 \\
DeepSeek & 0.889 & 0.\textbf{963} & 0.927 & 0.932 & 0.932 \\
aiXapply-RL & 0.89 & 0.958 & 0.936 & 0.958 & 0.938 \\
aiXapply-SFT & \textbf{0.924} & 0.937 & \textbf{0.946} & \textbf{0.958} & \textbf{0.941} \\
\bottomrule
\end{tabular}
\end{table}

\begin{table*}[!t]
\centering
\caption{Accuracy under randomized placeholders and chunked-file editing.}
\label{tab:combined-placeholder-chunk}
\scriptsize
\setlength{\tabcolsep}{3pt}

\resizebox{\textwidth}{!}{%
\begin{tabular}{lcccccccccccccccccccccc}
\toprule
\multirow{2}{*}{\textbf{Model}} 
& \multicolumn{11}{c}{\textbf{Random Placeholder}} 
& \multicolumn{11}{c}{\textbf{Chunk File}} \\
\cmidrule(lr){2-12} \cmidrule(lr){13-23}
& C & C++ & Docker & Go & HTML & Java & JS & JSON & Python & XML & Avg
& C & C++ & Docker & Go & HTML & Java & JS & JSON & Python & XML & Avg \\
\midrule

Qwen3-4B
& 0.6615 & 0.6328 & 0.76 & 0.675 & 0.60 & 0.715 & 0.68 & 0.7963 & 0.7128 & 0.70 & 0.696
& 0.5692 & 0.625 & 0.64 & 0.575 & 0.46 & 0.385 & 0.565 & 0.4815 & 0.6308 & 0.32 & 0.527 \\

DeepSeek 
& 0.9462 & 0.9297 & 0.98 & \textbf{0.975} & 0.90 & \textbf{0.945} & 0.9397 & 0.9259 & 0.9179 & 0.88 & 0.932
& 0.885 & 0.898 & 0.90 & 0.912 & 0.80 & 0.885 & 0.885 & 0.796 & 0.923 & 0.60 & 0.850 \\

RL 
& \textbf{0.9538} & 0.9375 & \textbf{1.00} & 0.925 & 0.90 & 0.935 & 0.955 & \textbf{1.00} & 0.9179 & \textbf{0.96} & 0.948
& 0.892 & 0.906 & \textbf{1.00} & 0.863 & 0.80 & 0.915 & 0.91 & 0.815 & 0.862 & 0.84 & 0.881 \\

SFT 
& 0.939 & \textbf{0.945} & \textbf{1.00} & 0.913 & \textbf{0.980} & 0.930 & \textbf{0.965} & \textbf{1.00} & \textbf{0.944} & 0.94 & \textbf{0.951}
& \textbf{0.908} & \textbf{0.922} & 0.92 & \textbf{0.875} & \textbf{0.88} & 0.910 & \textbf{0.935} & \textbf{1.00} & \textbf{0.887} & 0.80 & \textbf{0.900} \\

\bottomrule
\end{tabular}%
}

\end{table*}

\begin{table}[!t]
\centering
\caption{Error type distribution (\%) among classified errors across standard and long-context settings.}
\label{tab:error-analysis-merged}

\small
\setlength{\tabcolsep}{4pt}

\resizebox{\columnwidth}{!}{%
\begin{tabular}{lcccccc}
\toprule
& \multicolumn{4}{c}{\textbf{Standard}} & \multicolumn{2}{c}{\textbf{Long Context}} \\
\cmidrule(lr){2-5} \cmidrule(lr){6-7}
\textbf{Error Type} & \textbf{Qwen} & \textbf{DeepSeek} & \textbf{SFT} & \textbf{RL} & \textbf{SFT} & \textbf{RL} \\
\midrule
Output Invalid & 12.35 & 10.22 & 3.30 & 4.76 & 0.00 & 0.00 \\
Patch Not Applied & 8.29 & 5.11 & 8.79 & 9.52 & 0.00 & 0.00 \\
Patch Incomplete & 12.87 & 18.98 & 30.77 & 27.61 & 0.00 & 6.25 \\
Patch Incorrect & 4.06 & 8.03 & 9.89 & 11.43 & 12.50 & 0.00 \\
Wrong Position & 5.64 & 3.65 & 5.49 & 10.48 & 0.00 & 6.25 \\
Out-of-Patch Side Effect & 56.79 & 54.01 & 41.75 & \textbf{36.19} & 87.50 & 87.50 \\
\bottomrule
\end{tabular}%
}

\end{table}

\subsection{RQ2: Effectiveness Across Languages and Generalization}

Table~\ref{tab:main-benchmark} reports the main multi-language results. Task-specific training substantially improves a 4B model. The untrained 4B baseline achieves 0.651 average accuracy, while strong large-model baselines such as DeepSeek-V3.2, GLM-5, and Qwen3.5-397B reach 0.916, 0.921, and 0.948, respectively. 

Our task-specialized 4B models achieve 0.938 (\textbf{aiXapply-RL}) and 0.944 (\textbf{aiXapply-SFT}), demonstrating that a small model can close most of the gap to large models and approach state-of-the-art performance on this task. Notably, the SFT model performs competitively with the strongest baseline (0.944 vs.\ 0.948), while significantly outperforming both the untrained baseline and prior Apply-specific models.

Across the evaluated languages, both aiXapply variants are competitive with large-model baselines under the language-aware comparison. These results support transfer within the tested settings, rather than general language independence or repository-disjoint generalization.

\paragraph{Generalization} To evaluate whether the model truly learns the \textit{Apply} task—rather than overfitting to a fixed prompt format—we construct four out-of-distribution (OOD) settings that stress different aspects of robustness: long-context reasoning, placeholder variability, cross-language transfer, and partial-file editing.

\begin{itemize}
\item \textbf{Long Context:} Inputs exceed 32k tokens. Models often fail by truncating outputs or losing global consistency. This setting tests whether the model can preserve full-file integrity under long-context conditions.
\item \textbf{Randomized Placeholders:} We vary placeholder patterns (e.g., \texttt{// ... existing code ...}) into diverse formats. This evaluates whether the model relies on superficial pattern matching or truly understands structural insertion points.
\item \textbf{Untrained Languages:} We evaluate on languages not included in Apply training (e.g., CSharp, CSS, PHP, SystemVerilog) to measure cross-language generalization.
\item \textbf{Chunk File:} Instead of full files, inputs are truncated to syntactic units (e.g., functions or classes) or contiguous code spans. This simulates IDE scenarios where users select a code region and apply modifications locally.
\end{itemize}

Long-context editing is the most challenging setting among all OOD evaluations (Table~\ref{tab:long-context}). All models exhibit significant performance degradation, with the baseline collapsing to 0.2353, indicating severe failure in maintaining global consistency. While DeepSeek-V3.2 drops to 0.588, the RL-based model shows only limited gains (0.6471), suggesting that it struggles to preserve long-range structure. In contrast, \textbf{aiXapply-SFT} maintains a substantially higher accuracy (0.843), indicating stronger robustness in long-context scenarios.

As shown in Table~\ref{tab:combined-placeholder-chunk}, both apply models are robust to placeholder variations, in some scenarios, there are even performance improvements (Deepseek V3.2-C ,C++). This indicates that the models learn structural alignment well rather than relying on fixed placeholder tokens.

Results in Table~\ref{tab:untrained} show that the aiXapply models maintain strong performance on languages not seen during training, remaining competitive with a strong large-model baseline. The performance gap between seen and unseen languages is small, indicating that the model does not rely heavily on language-specific patterns. This suggests that the Apply task is largely language-agnostic and primarily depends on structural editing behaviors—such as locating the target region and preserving unchanged content—rather than detailed knowledge of a particular programming language.

Chunk-level editing is substantially more challenging than the full-file setting. Compared with the main benchmark (Table~\ref{tab:main-benchmark}), all models experience a noticeable drop in accuracy (Table~\ref{tab:combined-placeholder-chunk}), indicating that removing global context makes localization and integration more difficult. This degradation is consistent across models, but the impact varies. The baseline drops sharply, while DeepSeek-V3.2 also shows a clear decline. In contrast, the Apply models remain relatively stable and significantly outperform the baseline. Notably, \textbf{aiXapply-SFT} still achieves an average accuracy of 0.900, despite the reduced context. This suggests that the model has learned a robust local integration behavior, allowing it to correctly apply edits even when only partial code is available.

\paragraph{Error Analysis.}
We analyze model failures across both standard and long-context settings. Table~\ref{tab:error-analysis-merged} summarizes the error distributions.

\textbf{Overall patterns.} On the aix-apply test set, both SFT and RL models exhibit similar error distributions, with failures primarily concentrated in \textsc{OutOfPatchSideEffect} and \textsc{PatchIncomplete}. This indicates that the core challenge of the Apply task lies in maintaining strict locality while ensuring full patch application.

\textbf{Effect of training paradigm.}
Compared to baseline models (Qwen and DeepSeek), both SFT and RL have lower shares of \textsc{OutOfPatchSideEffect} among classified errors. In particular, RL achieves the lowest share (36.19\%), which does not alone establish fewer side effects or a causal effect of our error-aware reward design that explicitly penalizes unintended edits outside the target region. By assigning higher penalties to side effects, RL is encouraged to enforce strict locality and avoid out-of-patch modifications.

These differences lead to distinct editing behaviors. While both SFT and RL show relatively high \textsc{PatchIncomplete} shares (30.77\% and 27.61\%), these conditional percentages do not establish that RL produces fewer incomplete edits. RL also shows higher shares of \textsc{WrongPosition} and \textsc{PatchIncorrect}, suggesting a more aggressive editing strategy under ambiguous contexts. Overall, RL is designed to prioritize correct and localized edits, reflecting the optimization of error-aware rewards, whereas SFT favors stable token-level alignment with the target output.

\textbf{Long-context degradation.}
In long-context scenarios, the error distribution shifts dramatically and becomes highly concentrated, with both SFT and RL overwhelmingly dominated by \textsc{OutOfPatchSideEffect} (87.50\%). This indicates that preserving strict locality is the primary challenge in large contexts. Compared to the standard setting, the differences between SFT and RL largely diminish in error composition, and both models exhibit similar error patterns, suggesting that long-context editing introduces a common failure mode that dominates behavior regardless of the training paradigm.

\begin{tcolorbox}[ colback=gray!15, colframe=white, boxrule=0pt, arc=4pt, left=6pt,right=6pt,top=6pt,bottom=6pt ]
\textbf{Summary:} Except for very long code files, aiXapply maintains stable, high accuracy across the OOD settings (Tables~\ref{tab:long-context}--\ref{tab:combined-placeholder-chunk}). Performance under placeholder variation and cross-language transfer indicates that the model learns structural integration patterns rather than superficial formatting. Chunk-level editing remains effective, while very long files remain a limitation, particularly for RL.
\end{tcolorbox}

\subsection{RQ3: Efficiency and Cost}

A central question is whether full-file Apply—despite requiring full-file generation—can meet practical latency and cost constraints. Table~\ref{tab:rq3_main} compares decoding strategies on 165 examples sampled by language from the main benchmark (seed=42), whereas Table~\ref{tab:rq1_efficiency} uses all 1,637 examples.

\begin{table}[t]
\centering
\caption{Key efficiency results across decoding strategies.}
\small
\resizebox{\columnwidth}{!}{
\begin{tabular}{lccc}
\toprule
Method & Avg Latency (s) & P95 (s) & Throughput (tokens/s) \\
\midrule
Baseline (no speculation) & 28.83 & 90.23 & 102.04 \\
Suffix (default) & 5.75 & 20.74 & 509.54 \\
N-gram (default) & 2.17 & 6.94 & 1343.99 \\
N-gram (best, n=7, k=128) & \textbf{1.06} & \textbf{3.38} & \textbf{2692.01} \\
\bottomrule
\end{tabular}
}

\label{tab:rq3_main}
\end{table}

Table~\ref{tab:rq3_main} shows that speculative decoding fundamentally changes the efficiency profile of full-file Apply. Compared to the baseline (28.83s average latency), suffix decoding reduces latency to 5.75s, while n-gram further reduces it to 2.17s. The best configuration (n=7, k=128) achieves 1.06s latency and 2692.01 tokens/s throughput, corresponding to an overall $\sim$26$\times$ speedup over baseline.

\paragraph{Why n-gram dominates.}
Even under optimal tuning, suffix decoding remains significantly slower than n-gram, with a $\sim$2.6$\times$ gap in both latency and throughput. This difference aligns with the task structure: full-file Apply requires copying long spans from the input, and n-gram directly exploits exact substring reuse, making it better suited than suffix-based continuation heuristics.

\paragraph{Absolute efficiency.}
On the decoding subset, the best configuration achieves 1.06s mean latency, 3.38s P95, and 2692.01 tokens/s; the full-benchmark mean is 1.45s with 3734 output tokens per request. Together with the small model, these decoding gains support interactive deployment.

\begin{tcolorbox}[
    colback=gray!15,   %
    colframe=white,    %
    boxrule=0pt,       %
    arc=4pt,           %
    left=6pt,right=6pt,top=6pt,bottom=6pt
]
\textbf{Summary:} RQ3 is answered affirmatively, although full-file Apply involves longer outputs, it can be made efficient in practice through task-aligned model design and speculative decoding. The proposed approach achieves substantial reductions in latency and cost while maintaining strong performance, making full-file Apply viable for interactive use.
\end{tcolorbox}

\subsection{RQ4: Training Paradigm}

To better understand the differences between training paradigms, we further evaluate cross-format generalization on the \textit{Diff-XYZ} benchmark, which requires models to apply \emph{unified diff} inputs to source files. Compared with full-file Apply, this setting is more sensitive to formatting and alignment errors, since models must correctly interpret hunk headers, context lines, and line-level changes.

\begin{table}[!t]
\centering
\small
\caption{Performance on Diff-XYZ benchmark.}
\label{tab:diff-xyz}
\resizebox{\columnwidth}{!}{
\begin{tabular}{lcccccc}
\toprule
\textbf{Model} & \textbf{Overall} & \textbf{Java} & \textbf{JS} & \textbf{Kotlin} & \textbf{Python} & \textbf{Rust} \\
\midrule
Qwen3-4B & 0.596 & 0.630 & 0.610 & 0.620 & 0.590 & 0.530 \\
aiXapply-SFT     & 0.479 & 0.510 & 0.470 & 0.490 & 0.425 & 0.500 \\
aiXapply-RL      & \textbf{0.664} & \textbf{0.680} & \textbf{0.715} & \textbf{0.655} & \textbf{0.645} & \textbf{0.625} \\
\bottomrule
\end{tabular}
}
\end{table}

As shown in Table~\ref{tab:diff-xyz}, aiXapply-RL achieves the best performance on Diff-XYZ, outperforming both Qwen3-4B and aiXapply-SFT overall and across every language. This suggests that RL improves robustness to alternative edit representations even without explicit diff-format training, whereas SFT specializes more strongly to the full-file Apply format.

This contrasts with the full-file Apply results. On the main benchmark (Table~\ref{tab:main-benchmark}), SFT slightly outperforms RL (0.944 vs.\ 0.938). The same pattern holds across Random Placeholders and Chunk File (Table~\ref{tab:combined-placeholder-chunk}) and Untrained Languages (Table~\ref{tab:untrained}). The largest difference appears in Long Context (Table~\ref{tab:long-context}), where SFT is substantially stronger than RL (0.843 vs.\ 0.6471), indicating better preservation of global file structure. Overall, these results suggest different inductive biases: SFT provides stronger token-level fidelity and structural preservation within the full-file Apply paradigm, while RL offers better flexibility under distribution shifts such as diff-based editing.

\paragraph{Error distribution comparison.}
The error distributions in Table~\ref{tab:error-analysis-merged} complement these results. Both models are dominated by out-of-patch side effects and incomplete edits, and both reach an 87.50\% side-effect share among long-context errors. These conditional shares do not establish absolute error reductions. SFT's token-level supervision may help preserve unchanged regions, whereas RL's outcome-level feedback may allow greater flexibility across edit representations.

\begin{tcolorbox}[
    colback=gray!15,
    colframe=white,
    boxrule=0pt,
    arc=4pt,
    left=6pt,right=6pt,top=6pt,bottom=6pt
]
\textbf{Summary:} SFT and RL exhibit complementary strengths. SFT is more effective for in-distribution performance and long-context robustness, benefiting from token-level supervision and precise structural preservation. In contrast, RL demonstrates stronger cross-format generalization, suggesting better adaptability to alternative editing representations. These results indicate that the choice of training paradigm should depend on the target deployment setting, as neither approach is universally superior.
\end{tcolorbox}

\section{Industrial Deployment and Lessons Learned}

\textbf{Deployment setup.}
We further deployed aiXapply in the aiXcoder \cite{li2025aixcoder} IDE plugin as a dedicated Apply service. In this deployment, the upstream large model first generates an update snippet, and aiXapply then produces the full updated file, which is rendered as an editor diff for user inspection before acceptance. The previous workflow used the upstream large model itself to perform full-file Apply. In contrast, aiXapply-4B-RL is served on a single A100 40GB GPU and is invoked only for the Apply stage, without replacing the main planning or code-generation model. We selected RL following internal validation that favored RL over SFT on proprietary naming conventions, code structures, and project domains; this observation is specific to the deployment setting.
We reused otherwise-idle A100 capacity because the available H-series GPUs were allocated to large-model serving.

\textbf{Operational safeguards.}
Very large single files are rare in production. Training includes no-op examples, allowing the model to predict an empty edit and leave the source file unchanged. Developers can also undo Apply in the IDE and switch to the upstream model.

\textbf{Production latency.}
We compared end-to-end workflow latency before and after enabling aiXapply-RL over a one-week window in an enterprise deployment serving 200 internal users and averaging 8,000 model API calls per hour, as recorded by the New API middleware. Timing excluded queuing and included multiple rounds of tool calls. Before deploying aiXapply, an average IDE editing task took about 240 seconds, where the Apply stage contributed 50 seconds on average and 89 seconds at P95. After offloading Apply to aiXapply, the Apply latency drops to 1.89 seconds on average and 3.78 seconds at P95, corresponding to a 26.5$\times$ reduction in mean Apply latency and a 23.5$\times$ reduction in P95 Apply latency. We cannot report a precise measured post-deployment end-to-end mean. Holding the other workflow stages fixed gives an estimated average end-to-end latency of $240 - 50 + 1.89 = 191.89$ seconds.

\textbf{Serving capacity and cost.}
Offloading full-file generation motivates a capacity-planning scenario. If the estimated reduction in workflow time translates proportionally into capacity under comparable workload and utilization, five upstream 8$\times$H200 serving instances of a 600B model could be replaced by four, plus one shared A100 40GB aiXapply service. One aiXapply instance can serve the Apply workload generated by approximately eight upstream IDE model services, so the additional Apply cost is amortized across multiple large-model services. In normalized machine-cost terms, this scenario changes $5C_{\mathrm{8H200}}$ to $4C_{\mathrm{8H200}} + C_{\mathrm{A100}}$: a 20\% reduction in upstream machine cost before the A100 cost. Workflow elapsed time is not GPU service time; this scenario requires capacity validation and cost allocation, rather than establishing a measured net saving.

\textbf{Human evaluation in real development tasks.}
We evaluated the Apply module in 100 development tasks across approximately 40 internal aiXcoder projects, mainly in TypeScript, Java, and Python. Each task concerned a feature or bug fix; we evaluated one Apply invocation per task. Tasks were selected from accumulated project requirements, mainly in backend development, for their clear descriptions and representativeness. The update snippets were generated by GLM-5 during normal development, rather than synthesized specifically for evaluation. Within these tasks, aiXapply-RL and the upstream GLM-5 model performed Apply using the same source files and update snippets. We evaluated only the first outputs, without retries or manual repairs. The 100 cases were divided among five testing-team members, with one assessor per case, and their results were aggregated. Acceptance required applying the upstream-generated snippet at the correct location, preserving unrelated source content, and faithfully integrating its core logic without semantic errors. Apply was judged acceptable in 91 of 100 cases for aiXapply-RL and 88 of 100 for GLM-5. These are descriptive single-Apply acceptance rates, not end-to-end task success rates. Following this internal evaluation, the RL service was rolled out to aiXcoder enterprise users, reaching thousands of users.

\textbf{Open-source integration.}
To make the deployment reusable, our open-source artifact provides a Continue adapter and an OpenAI-compatible endpoint, supporting both local and remote deployment. Developers can therefore keep the original upstream model for planning and snippet generation while routing Apply requests to aiXapply as a specialized backend.
The same service can also be exposed as an agent tool, providing an alternative to search-and-replace for snippet integration in broader agent workflows, with the aim of improving accuracy and reducing latency.

\textbf{Lessons learned.}
This deployment illustrates three practical choices: separate snippet generation from copy-dominant integration; select the Apply model using internal project validation as well as public benchmarks; and retain the editor diff for user inspection before acceptance. The observed latency gains and single-Apply manual assessments support this deployment, while broader agent use remains an integration opportunity rather than an evaluated outcome.

\section{Threats to Validity}

\textbf{Internal validity.}
We compare SFT and RL under the same backbone and evaluation protocol, but differences in optimization and reward design may still affect results. RL may also introduce variance from reward quality and sampling.

\textbf{External validity.}
Our experiments use the aiXapply and Diff-XYZ datasets, covering multiple languages, structured-text formats, and OOD settings. However, they may not represent all real-world codebases or interactive IDE workflows, so generalization to all deployment settings is not guaranteed.
The shared snippet-generation pipeline may introduce synthesis bias, and LLM verification may accept underspecified snippets. The commit-level random split does not guarantee repository-disjoint generalization. The real-development evaluation provides complementary evidence of practical applicability, but does not eliminate this bias. Our training experiments use only Qwen3-4B, so transfer to other backbones remains unverified. Evaluation is limited to single-file Apply; multi-file workflows could invoke it per file, but cross-file coordination must be handled by the surrounding agent.

\textbf{Conclusion validity.}
The main trends are consistent across settings, but some subsets, such as long-context evaluation, are relatively small. Thus, the exact magnitude of differences between SFT and RL should be interpreted cautiously.

\section{Conclusion}

We study snippet integration as a standalone task: merging an update snippet into a source file while preserving all other content unchanged.
We analyze common editing paradigms, including unified diff and search-and-replace, and show that their patch-structure and source-matching constraints lead to unreliable one-shot performance.

We propose \textbf{aiXapply}, a specialized approach that directly generates the full updated file.
By constructing a dataset from real commits and training a 4B model with supervised fine-tuning and reinforcement learning, we obtain a system that is designed to produce deterministic and side-effect–free edits.

The benchmarks show competitive reference-agreement accuracy, while the aiXcoder deployment demonstrates lower Apply latency and acceptable first outputs in the evaluated internal tasks.

More broadly, our findings suggest that code editing should be treated as a distinct and well-defined problem rather than a byproduct of general code generation.
Specializing models for such tasks enables a better balance between accuracy, efficiency, and deployability.

In future work, we plan to extend Apply to more complex editing scenarios, such as multi-file changes, interactive editing workflows, and tighter integration with development tools.

\section{Data Availability Statement}
The code is available in the \url{https://github.com/aixcoder-plugin/aiXapply-4B} repository. The \href{https://huggingface.co/datasets/aiXcoder/aiXapply_test_data}{test set} and \href{https://huggingface.co/aiXcoder/aiXapply-4B-SFT}{SFT}/\href{https://huggingface.co/aiXcoder/aiXapply-4B-RL}{RL} weights are available on Hugging Face. 

\newpage
\bibliographystyle{IEEEtran}
\bibliography{apply-base}

\end{document}